\documentclass[wssquare]{ws-procs961x669}

\pdfoutput=1
\usepackage{graphicx}
\usepackage{xcolor}
\usepackage[caption=false]{subfig} 
\usepackage{amssymb}
\usepackage{tensor}
\usepackage{adjustbox}
\usepackage{booktabs}
\usepackage{tabularx}
\usepackage{multirow}
\usepackage{enumitem}
\usepackage{etoolbox}

\usepackage[caption=false]{subfig}
\newcommand{\subfigimg}[3][,]{%
	\setbox1=\hbox{\includegraphics[#1]{#3}}%
	\leavevmode\rlap{\usebox1}%
	\rlap{\hspace*{-18pt}\raisebox{.5\baselineskip}{\small{#2}}}%
	\phantom{\usebox1}%
}
\usepackage{url,hyperref}
\hypersetup{colorlinks,linkcolor={blue!55!black},citecolor={red!45!black},urlcolor={blue!45!black},breaklinks=true}

\begin{document}

\title{Properties of dynamical regular black holes in semiclassical gravity}

\author{Ioannis Soranidis}

\address{School of Mathematical and Physical Sciences, Macquarie University, \\Sydney, New South Wales 2109, Australia \\E-mail: \href{mailto:ioannis.soranidis@hdr.mq.edu.au}{ioannis.soranidis@hdr.mq.edu.au}}

\begin{abstract}
Regular black holes have become a popular alternative to the singular mathematical black holes predicted by general relativity as they circumvent mathematical pathologies associated with the singularity while preserving crucial black hole features such as the trapping of light. Based on the assumption that semiclassical gravity is valid in the vicinity of their apparent horizons, we review their thermodynamic properties in static scenarios and demonstrate that 1-loop corrections to the Bekenstein-Hawking entropy, accounting for the entanglement entropy of fields, depend on the regularization scheme used. Taking backreaction effects into consideration, we investigate their full dynamical evolution and we examine the behavior of the null energy condition. We find that it is always violated in the vicinity of the outer horizon while being satisfied in the vicinity of the inner horizon, which implies that the trapped spacetime region, as determined from the behavior of null geodesic congruences, is effectively separated into a null energy condition-violating and a non-violating domain. Our findings suggest that quantum effects are more dominant close to the outer apparent horizon and become more pronounced towards the final stages of the evaporation process.

\end{abstract}

\keywords{black holes; semiclassical gravity; thermodynamics; energy conditions; quantum aspects of black holes; entanglement entropy.\\[3mm]
Contribution to the proceedings of the 17th Marcel Grossmann meeting (7--12 July 2024).\\
Session classification: Wormholes, energy conditions, and time machines}

\bodymatter
\thispagestyle{empty}

\section{Introduction}\label{sec:introduction}

In 1965, Penrose demonstrated that under the assumptions of global hyperbolicity, the existence of a noncompact Cauchy hypersurface, the validity of classical Einstein equations, and the non-violation of the null energy condition (NEC), a spacetime containing trapped surfaces is inevitably geodesically incomplete, introducing the first ``modern" singularity theorem \cite{P:65}. The existence of inextensible geodesics is commonly equated with the presence of regions of unbounded curvature, known as gravitational singularities. While we adopt this assumption in this article, it is crucial to note that this equivalence does not always hold true \cite{W:20,SW:17, SW:17-essay}. Singularities pose a significant challenge as they signal the breakdown of the underlying theory, fundamentally limiting our ability to predict the future  \cite{H:76}. \textit{``When a theory predicts singularities the theory is wrong}!" \cite{K:23} or undoubtedly incomplete. It is anticipated that the merging of general relativity with quantum field theory into a unified theory of quantum gravity will resolve the singularity issues. Quantum effects are expected to play a pivotal and indispensable role in this resolution.

In the absence of a complete theory, research efforts are partially directed towards effective models capable of mimicking observable phenomena, such as light rings, while circumventing singularity issues. These models, known as black hole mimickers, are categorized into two distinct groups based on the absence or presence of a horizon \cite{CP:19}. The former corresponds to horizonless ultracompact objects, which have the capacity to form two light rings around them, with one being stable and is linked to nonlinear spacetime instabilities \cite{CBH:17,DF:24}. This property potentially renders them undesirable candidates for the objects we currently observe. However, it has been  shown that horizonless ultracompact objects sourced by nonlinear electrodynamics (NED) can admit only one unstable light ring due to the phenomenon of birefringence \cite{MS:24}. The latter category comprises regular black holes (RBHs), which are trapped spacetime regions bounded by both an inner and an outer horizon and they are the main focus of this article.  

Current astrophysical observations, although in close agreement with the Kerr metric, are unable to distinguish between black hole mimickers and genuine black holes with singularities. Therefore, it is premature to conclusively identify the observed ultracompact objects as black holes \cite{M:23}. Furthermore, asserting that these objects are indeed black holes requires the observation of the event horizon. However, such an entity is inherently teleological and fundamentally unobservable, even in principle \cite{V:14}. It could be argued that the detection of a quasilocal apparent horizon should suffice to verify the presence of a trapped region but from the currently available data its existence cannot be inferred. Unfortunately, neither category of black hole mimickers constitutes a panacea to all of our problems, as substituting the singularity with an effectively regular spacetime results in peculiar features.

All singularity theorems rely on the assumption that collapsing matter behaves in accordance with certain energy conditions, which describe classical matter and aim to capture the property of the non-negativity of mass and the attractive nature of gravity \cite{S:98,SG:15}. However, quantum fields systematically violate these classical energy conditions and consequently influence the theorems of general relativity in various ways \cite{KS:20}. For instance, during gravitational collapse, when matter is compressed to high densities, quantum effects are expected to become significant and violation of energy conditions may occur. But how does such a violation contribute to the resolution of singularities and the emergence of regular black hole spacetimes?

A star with appropriate mass will undergo gravitational collapse once sufficient thermal energy has been radiated away. The first mathematical model describing complete gravitational collapse was formulated by Oppenheimer and Snyder \cite{OS:39}. In this model, a spherically symmetric, homogeneous, collisionless matter collapses into a Schwarzschild black hole under the influence of gravity's attractive force. The formation of the singularity can only be prevented by introducing some repulsive effect to halt the collapse. This effect may arise from the breakdown of general relativity during the final stages of collapse and/or from violation of the NEC \cite{Bambi:book:23}. Various horizonful or horizonless geometries can emerge from this collapse, including RBHs, black-to-white hole bounces, and horizonless remnants \cite{MT:22, BMM:13}. 

\section{Minimal length scale, thermodynamics, and entanglement entropy}\label{sec:static}
The introduction of a minimal length scale $\ell$ is a convenient way to describe the spacetime geometry after a ``regular" collapse. It is important to note that while this length is presumed to arise from a quantum gravity theory and signifies a length scale where modifications of Einstein equations become significant \cite{F:16}, it is distinct from the Planck length, $\ell_p$ \cite{COS:22,CLMMOS:23}. This characteristic is common among all the regular geometries mentioned in Sec.~\ref{sec:introduction}. To elucidate some of their properties, we use three well-established models in the literature: the Hayward model \cite{H:06}, the Bardeen model \cite{B:68}, and the model proposed in Ref.~\cite{CLMMOS:23} that we will refer to as Cadoni \textit{et al.} model. The line element we use to describe the aforementioned models has the following form:
\begin{align}
	ds^2=-f_{i}(r)dv^2+2dvdr+r^2d\Omega_{2}, \label{eq:ds}
\end{align}     
where $v$ is the advanced coordinate, $r$ is the areal radius, $d\Omega_{2}=d\theta^2+\sin^2{\theta}~d\phi^2$ denotes the line element of the
2-sphere, $i=\mathcal{H}, \mathcal{B}, \mathcal{C}$ represents the model we are considering and the relevant metric functions are 
\begin{align}
	f_{\mathcal{H}}(r)=1-\frac{2mr^2}{r^3+2m\ell^2},\quad f_{\mathcal{B}}(r)=1-\frac{2mr^2}{(r^2+\ell^2)^{3/2}}, \quad f_{\mathcal{C}}(r)=1-\frac{2mr^2}{(r+\ell)^3}.\label{eq:models}
\end{align} 

\begin{figure}[!htbp]
	\hspace*{1.8cm}
	\includegraphics[scale=0.5]{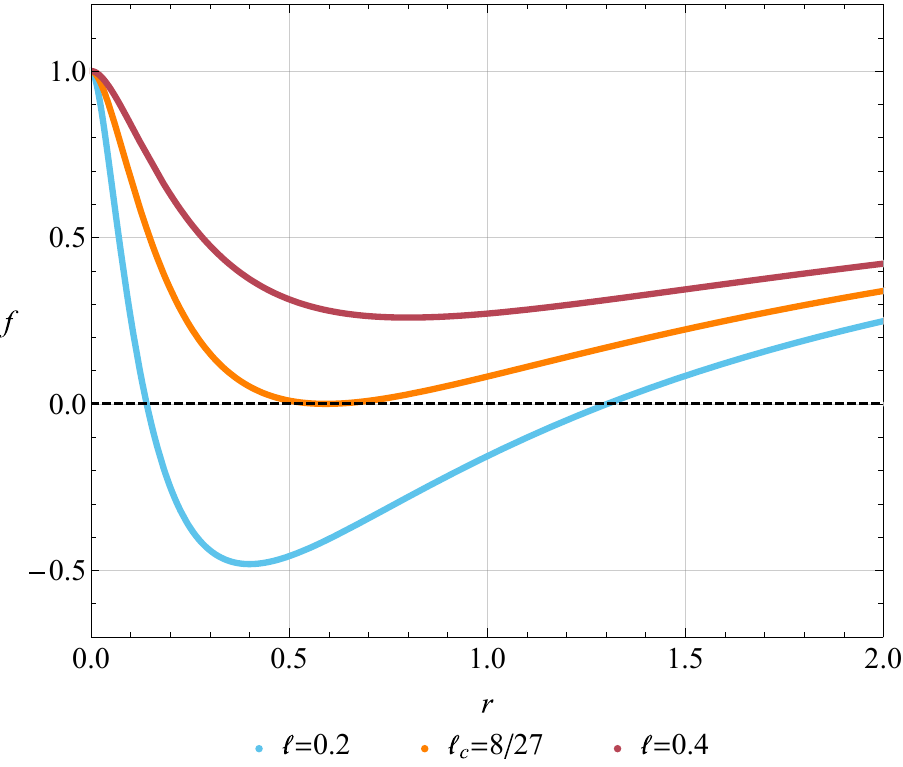}
	\caption{The metric function for the Cadoni \textit{et al.} model, $f_{\mathcal{C}}(r)$, is depicted for $m=1$ for varying minimal length scales $\ell$. The critical length, $\ell_{c}=8/27$, is indicated by the orange line, representing an extremal RBH. For $\ell<\ell_{c}$, the light blue line represents an RBH with two horizons that bound the trapped region of spacetime, whereas for $\ell>\ell_{c}$, the red line illustrates a horizonless configuration.} 
	\label{fig:RBH-UCO}
\end{figure}

In all of these models, the introduction of the minimal length scale $\ell$ gives rise to a de Sitter or anti-de Sitter (AdS) core, which in turn leads to finite curvature scalars at the center \cite{Bambi:book:23}. However, these models differ in the way they achieve singularity regularization and in their asymptotic behavior, exhibiting deformations of different strength from the Schwarzschild geometry, as follows:
\begin{align}
	f_{\mathcal{H}}(r)=1-\frac{2m}{r}+\frac{4m^2\ell^2}{r^4}+\mathcal{O}{(r^{-7})},\quad f_{\mathcal{B}}(r)=1-\frac{2m}{r}+\frac{3m\ell^2}{r^3}+\mathcal{O}{(r^{-5})},  \label{eq:asymp-behavior-1}
\end{align} 
\begin{align}
  f_{\mathcal{C}}(r)=1-\frac{2m}{r}+\frac{6m\ell}{r^2}+\mathcal{O}{(r^{-3})}. \label{eq:asymp-behavior-2}
\end{align}  
Depending on the value of the minimal length scale $\ell$, the geometries of Eq.~\eqref{eq:models} can represent either horizonful or horizonless objects. The spacetime will admit horizon(s) if the expansion of the outgoing light rays $\theta_{+}$ vanishes at a specific radius, or equivalently if the equation $f_{i}(r)=0$ admits real positive solutions \cite{H:94}. For all of these models there exists a critical length $\ell_{c}$ which determines what type of geometry is present: For $\ell<\ell_{c}$ the geometry is that of an RBH, for $\ell=\ell_{c}$ it corresponds to an extremal RBH, and for $\ell>\ell_{c}$ it describes a horizonless ultracompact object \cite{Bambi:book:23}. These types of geometries are illustrated in Fig.~\ref{fig:RBH-UCO} as light blue, orange, and red lines, respectively. In this article, we only consider the cases $\ell\leq \ell_{c}$ since we are concerned with the study of RBHs. 

A preliminary question is how to source such geometries in the absence of a quantum gravity theory from which they presumably arise. The response to this question is not unique, as quantum effects rendering the geometry regular can be simulated through various means. Two popular ways of achieving this is in four-dimensional spacetimes are through NED theories within the framework of general relativity \cite{FW:16} or through the scalar Einstein--Gauss--Bonnet theory \cite{NN:23}. For the analysis of thermodynamic properties we proceed with the former. NED theories have been successful in eliminating singularities linked with the self-energy of point charges \cite{BI:34}, but it was also demonstrated in Ref.~\cite{FW:16} that the inclusion of magnetic charge $Q_{m}$ suffices to cure gravitational singularities. For the metrics of Eq.~\eqref{eq:models} the minimal length scale is introduced via the magnetic charge through a relation $Q_{m}=Q_{m}(m,\ell)$, which vanishes in the case of $\ell=0$ \cite{FW:16,SS:24,S:24}. It is important to note that the asymptotic behavior described in Eqs.~\eqref{eq:asymp-behavior-1} and \eqref{eq:asymp-behavior-2} can be traced back to the weak-field limit of the respective NED Lagrangian used to generate these geometries \cite{MS:24}. Notably, the Cadoni \textit{et al.} model admits a Maxwell weak-field limit, which explains the $\mathcal{O}(r^{-2})$ corrections to the Schwarzschild asymptotic behavior, similar to that of the Reissner–Nordström (RN) black hole. In contrast, the Hayward and Bardeen models exhibit weak-field limits stronger than the Maxwell, resulting in weaker corrections of orders $\mathcal{O}(r^{-4})$ and $\mathcal{O}(r^{-3})$, respectively.

In the vicinity of black hole horizons, one of the most important quantum effects is Hawking radiation. Classically, black holes do not emit radiation because they are effectively at zero temperature. However, taking into account the behavior of quantum fields on a curved background, Hawking revealed that they can indeed radiate and possess a blackbody spectrum with a temperature directly proportional to the surface gravity \cite{H:75}. This radiation flux is observed at infinity and leads to the decrease of the black hole's mass. Additionally, black holes carry entropy that is intimately tied to their microscopic degrees of freedom. At the semiclassical level, neglecting backreaction effects, the entropy conforms to the Bekenstein--Hawking (BH) formula $A/4G$, where $A$ is the area of the black hole event horizon, and $G$ represents Newton's gravitational constant. These characteristics prompt the formulation of laws governing black hole thermodynamics in direct analogy with principles of conventional thermodynamics \cite{BCH:73}.

The first law of black hole mechanics describes the quasi-static transition between two equilibrium states of the black hole in terms of certain macroscopic quantities. In the extended phase space, for the RBH case, these thermodynamic quantities include the pressure/tension $P=-\Lambda/8\pi$ associated with the presence of a negative/positive cosmological constant $\Lambda$, the minimal length scale $\ell$ which is treated as a fundamental thermodynamic parameter to facilitate a theory-agnostic approach in describing the thermodynamic characteristics of the RBH models outlined in Eq.~\eqref{eq:models}, and the thermodynamic entropy $S$. The conjugate potentials to these thermodynamic quantities are the  temperature $T$, the potential $\Phi$, and the thermodynamic volume $V$, respectively. 

An extremely effective method for deriving these thermodynamic quantities in the semiclassical approximation is the Euclidean path integral formalism due to Gibbons and Hawking \cite{GH:77}. By finding the stationary point of the action or, equivalently, extremizing it with respect to the horizon radius, one can deduce the temperature $T$ and subsequently derive the remaining relevant thermodynamic quantities with the prescription of Ref.~\cite{BBWY:90}. Applying this methodology to the case of RBHs reveals that the form of the first law of black hole mechanics is given by 
\begin{align}
	dM=TdS+\Phi d\ell+VdP,
\end{align}
where $M$ is the Komar mass of the spacetime. It is worth noting that in a de Sitter spacetime achieving thermodynamic equilibrium poses a challenge due to the different temperatures of the black hole and cosmological horizons. To resolve this problem, one may introduce an isothermal cavity which in turn requires the introduction of an additional term $\lambda dA_{c}$ in the first law, where $\lambda$ is the conjugate potential of the cavity's area $A_{c}$ \cite{KS:16,SM:18,SM:19,HHMS:20,SFM:21,S-essay:23}. When treating the minimal length scale as a fundamental parameter, we observe the emergence of a modified temperature $T$, while the entropy retains the usual BH form \cite{SS:24,S:24}.

\begin{figure*}[!htbp]
	\centering
	\begin{tabular}{@{\hspace*{-0.0\linewidth}}p{0.57\linewidth}@{\hspace*{0.0\linewidth}}p{0.45\linewidth}@{}}
		\centering
		\subfigimg[scale=0.65]{(a)}{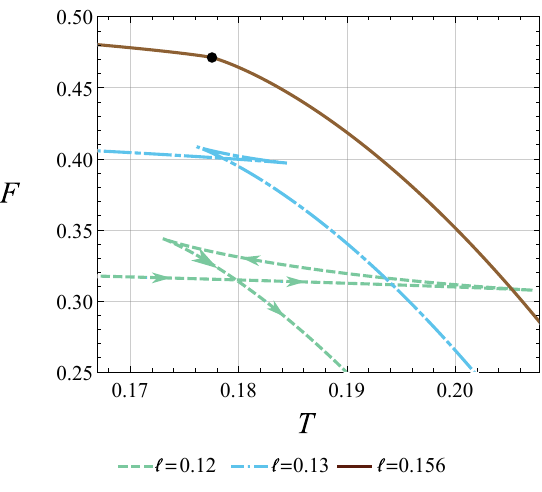} &
		\subfigimg[scale=0.58]{(b)}{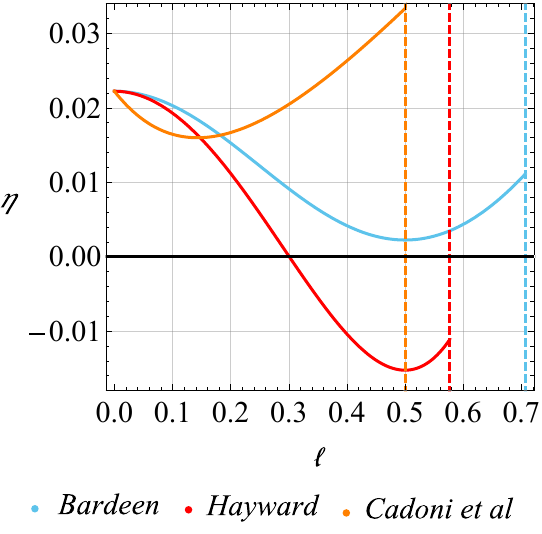} 
	\end{tabular}
	\caption{(a) Free energy $F=M-TS$ in the canonical ensemble as a function of the temperature $T$ for a constant value of the cosmological constant $\Lambda=-1$ (AdS spacetime) with varying minimal length $\ell$. The arrows indicate the direction of increasing horizon radius; (b) Coefficient of the logarithmic correction term $\eta$ for the entropy of the massless, minimally coupled, scalar field for three different RBH metrics embedded in Minkowski spacetime with respect to the minimal length scale $\ell$. The outer horizon radius is fixed at $r_+=1$. The vertical dashed lines represent the critical lengths $\ell_c$ for which the geometry corresponds to the extremal case for each RBH model.} \label{fig:PT-Sq}
\end{figure*}

In the extended phase space, one of the most intriguing phenomena is the occurrence of phase transitions \cite{KMT:17}. Embedded in AdS, the models described by Eq.~\eqref{eq:models} have common traits: they exhibit a characteristic ``swallowtail" behavior in the canonical ensemble. A small-to-large first-order phase transition occurs for varying minimal length scale $\ell$ which terminates with a second-order phase transition at a critical point indicated by a black dot in Fig.~\ref{fig:PT-Sq}(a). This behavior is analogous to the liquid--gas transition occurring in a traditional van der Walls fluid. In this particular fluid case and also in a large number of black holes, it is found that their critical point is described by the standard mean-field theory exponents and a critical ratio $P_{c}v_{c}/T_{c}=3/8$, where $P_c$, $T_c$, and $v_c$ are the pressure, temperature and reduced volume at the critical point, respectively \cite{KMT:17}. RBHs are no exception for the former statement but it is found that the stronger the deformations from the Schwarzschild geometry (as quantified by Eqs.~\eqref{eq:asymp-behavior-1} and \eqref{eq:asymp-behavior-2}) the larger the deviation from this ``universal" mean-field theory critical ratio \cite{S:24} signaling a departure from the typical van der Waals behavior. 

Furthermore, the absence of the Hawking--Page transition \cite{HP:83} is noteworthy and reminiscent of the RN case where the complete evaporation of the black hole is not possible. However, in the case of RBHs, this feature is inherently linked to the regularization of the spacetime. The Hawking--Page transition has a natural interpretation within the AdS/CFT correspondence, where the transition between a black hole and thermal radiation in the bulk is dual to a deconfinement transition in the boundary conformal field theory \cite{KMT:17}. The corresponding interpretation of the first-order phase transition in the bulk theory is a transition between a low-entropy and a high-entropy conformal field theory \cite{ACKMV:23}. 

The above calculation is at the semiclassical level, and additional corrections to the BH entropy can be found through the 1-loop effective action. We show that these corrections are contingent upon the value of the minimal length scale and the way the singularity is smoothed out. Following Refs.~\cite{S:94,S:11} we can explicitly compute the ``non-geometric" corrections to the BH entropy induced by the fluctuations of a scalar field on the background geometry. These corrections also represent the entanglement entropy of the scalar field. The explicit formula for four-dimensional black holes with a metric line element described by Eq.~\eqref{eq:ds}, for the case of a massless, minimally coupled, scalar field, is given by \cite{S:94,S:11}:
\begin{align}
	S_{q}=\frac{A}{48\pi \epsilon^2}+\eta \ln\left(\frac{r_+}{\epsilon}\right), \quad \eta=\frac{1}{18\pi}-\frac{A}{20\pi}\left(\frac{1}{6}f_{i}''(r_+)+\frac{f_{i}'(r_+)}{2r_+}\right).
\end{align} 
In Fig.~\ref{fig:PT-Sq}(b), we illustrate the coefficients $\eta$ for the three distinct models of Eq.~\eqref{eq:models}. It becomes evident that the choice of singularity regularization, and consequently the quantum effects it entails, directly influences the entanglement entropy of the fields. Specifically, when $\ell=0$, all geometries reduce to Schwarzschild, yielding $\eta=1/45$, a well-established value \cite{S:11}. Notably, the sign of $\eta$ remains positive for both the Bardeen and Cadoni \textit{et al.} models, whereas there is a sign change for the Hayward model. The latter behavior is similar to that observed in the RN case but stands in contrast with the other two models. These features underscore the impact of singularity regularization method on the entanglement entropy of the fields. 

We can consistently express the total entropy in the form of BH entropy by appropriately renormalizing Newton's constant and the horizon radius. Specifically, we can write 
\begin{align}
	S=S_{BH}+S_{q}=\frac{A}{4G}+\frac{A}{48\pi \epsilon^2}+\eta \ln\left(\frac{r_+}{\epsilon}\right)=\frac{\tilde{A}}{4\tilde{G}},
\end{align} 
where $\tilde{G}$ and $\tilde{A}=4\pi \tilde{r}^2_+$ are the quantum corrected expressions defined as:
\begin{align}
	\frac{1}{\tilde{G}}=\frac{1}{G}+\frac{1}{12\pi\epsilon^2}, \quad \tilde{r}^2_+=r^2_++\frac{\tilde{G}\eta}{\pi}\ln\left(\frac{r_+}{\epsilon}\right).
\end{align}
We observe that the quantum corrected radius depends on the sign of $\eta$. A positive sign implies that the quantum fluctuations of the massless scalar field induce the expansion of the radius, while a negative sign results in contraction. This provides a physical intuition as to how the presence of additional fields on the background geometry affects the horizon's position. It is worth noting that the horizon shifting is negligible for black holes of mass much greater than the Planck mass but on the contrary, at the Planck scale, these corrections become important \cite{S:94}. 

\section{Formation, evaporation, and ultimate fate}\label{sec:dynamic}
Up until now, our study neglected backreaction effects resulting from the presence of Hawking radiation. If we take them into account, we transition to a fully dynamical setting. As a result, in the line element described by Eq.~\eqref{eq:ds}, the metric function $f_i(r)$ becomes dependent on the advanced coordinate $v$ as well. This time evolution arises from the variation of mass $m=m(v)$ and the minimal length scale $\ell=\ell(v)$, which emerges due to the incorporation of backreaction effects \cite{BBCRG:21,BBCRG:22,CRFLV:23}. Our analysis is based on the following two assumptions: the validity of semiclassical gravity in the vicinity of both the inner horizon $r_{-}(v)$ and the outer horizon $r_{+}(v)$, and the finite formation of the RBH as perceived by a distant observer. Under these conditions, in spherically symmetric settings, only the scenario of shrinking future horizons remains permissible, as it is the only one that yields real-valued solutions to the Einstein equations \cite{MMT:22,DSST:23}. 
\begin{figure*}[!htbp]
	\centering
	\begin{tabular}{@{\hspace*{-0.04\linewidth}}p{0.57\linewidth}@{\hspace*{-0.055\linewidth}}p{0.55\linewidth}@{}}
		\centering
		\subfigimg[scale=0.35]{(a)}{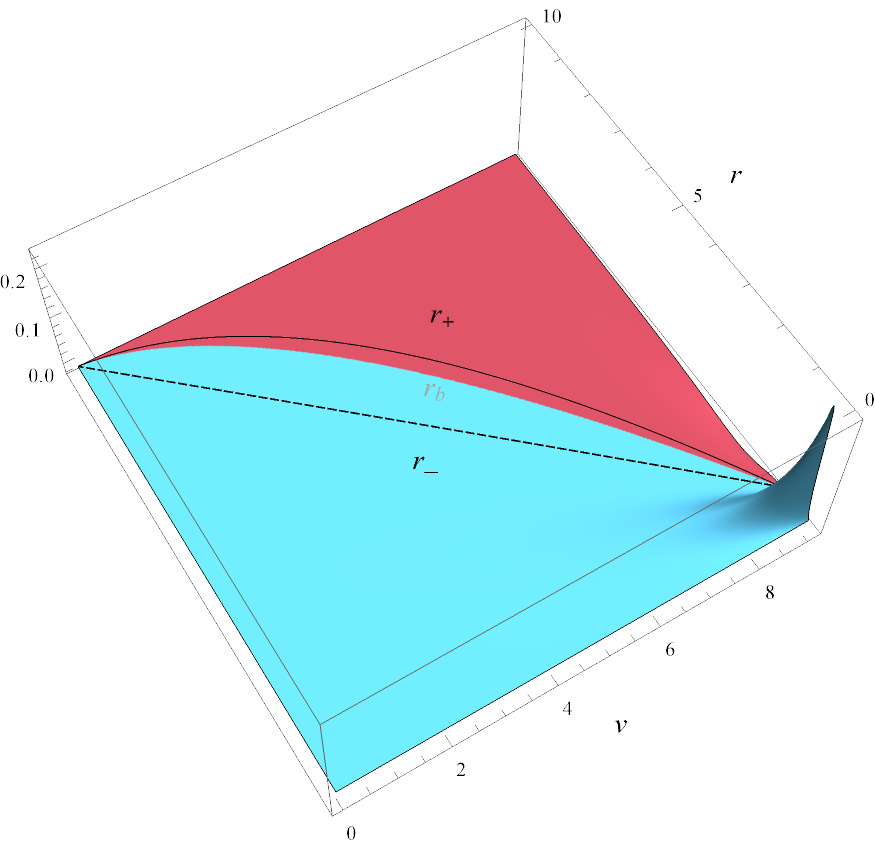} &
		\subfigimg[scale=0.3]{(b)}{NECevolution-2} 
	\end{tabular}
	\caption{Evaluation of the NEC for the entire evolution period $v \in \left[ v_\mathrm{f} , v_\mathrm{d} \right]$ of the evaporating, i.e., $r'_{\pm}(v)<0$, inner-extremal RBH proposed in Ref.~\cite{CDLPV:22}. The points $v=v_\mathrm{f}=0$ and $v=v_\mathrm{d}\approx8.8$ denote the advanced null coordinate at the formation and disappearance of the trapped region, respectively. (a) Status of the NEC within the trapped region and its vicinity. The spacetime domain where the NEC is violated (satisfied) is shaded in red (light blue). The hypersurface $r_{b}$ located between the two horizons corresponds to the boundary of the two domains. The behavior illustrated here is universal provided that  $r'_{\pm}(v)<0$ \cite{MS:23}; (b) The red (light blue) line represents the NEC evaluated at the outer (inner) apparent horizon $r_+(v)$ $\left(r_{-}(v)\right)$.} \label{fig:RBH}
\end{figure*}

The full dynamical evolution begins with the collapse of a sufficiently massive star. At a specific moment $v_{\mathrm{f}}$, the horizon forms, branching into inner and outer components which subsequently shrink. To maintain spacetime regularity, it is essential that the inner horizon never reaches the center $r=0$. This implies that there is a unique way for the evaporation to cease, namely through the re-merging of the inner and outer horizons. This state has a degenerate horizon but no trapped region, i.e., it is an extremal RBH. Similarly to the static scenario where the temperature is proportional to the surface gravity, in the dynamical case, the temperature is correlated with the Kodama surface gravity, which is the dynamical generalization of the Killing surface gravity \cite{K:80,AV:10,KPV:21}. For the extremal case, the Kodama surface gravity will vanish and one obtains a state of zero temperature, i.e., a cold remnant \cite{MS:23-therm}. 

While this may resolve the issue of maintaining regularity throughout the RBH evolution, it introduces another unfortunate situation: the violation of the third law of black hole thermodynamics. In this scenario a state of zero temperature can be achieved through a finite sequence of operations---a phenomenon widely believed to be impossible for real physical systems. Although the third law of ordinary thermodynamics can be violated in classical scenarios, such as in the case of an ideal gas, empirical evidence demonstrates its validity across all physical systems studied in laboratory settings \cite{W:97,R:00}. Therefore, quantum effects during the final stages of evaporation could potentially lead to a significant departure from our broader physical understanding of nature, possibly preventing the formation of such a cold remnant and rendering the extremal RBH an inaccesible limit in its dynamical evolution.

The violation of the third law is connected to the violation of the NEC \cite{I:86}, which represents the weakest of the energy conditions \cite{HE:73:book}. Studies have demonstrated its violation in the presence of Hawking radiation \cite{V:97,FN:98,W:01,LO:16}, making it interesting to explore its behavior over the course of the RBH's lifetime. Calculating the NEC for the metric of Eq.~\eqref{eq:ds} with dependence on the advanced coordinate $v$ and using the radial outgoing null vector $k^{\mu}=\left(1,f/2,0,0\right)$, we find that 
\begin{align}
	T_{\mu\nu}k^{\mu}k^{\nu}=-\frac{\partial_{v}f}{8\pi r}.
\end{align}
In Fig.~\ref{fig:RBH}(a) it is shown that the NEC is violated in the vicinity of the outer horizon and satisfied near the inner one. This results in an effective splitting of the trapped region into a NEC-violating and a non-violating region with the boundary represented by $r=r_{b}(v)$. Furthermore, as illustrated in Fig.~\ref{fig:RBH}(b), the evaluation of the NEC on the outer apparent horizon reveals that the maximal violation occurs towards the final stages of evaporation, implying that quantum effects are much more prominent there \cite{MS:23}. Finally, we should mention that the model of Ref.~\cite{CDLPV:22}, whose dynamical generalization is employed in the numerical analysis of the NED in this article, admits a degenerate inner horizon which aims to cure the mass inflation instability \cite{PI:89} and this property is associated with the vanishing of the NEC at the inner horizon, as illustrated by the light blue line in Fig.~\ref{fig:RBH}(b). It is worth noting that this model can arise as a solution to the vacuum gravity equations when considering an infinite series of quasi-topological higher curvature corrections in spacetimes with more than four dimensions \cite{DFKK:24}. Nonetheless, we stress that the behavior of other dynamical RBH models is qualitatively similar in the sense that the NEC is always violated near the outer horizon while being satisfied near the inner horizon, and quantum effects being more significant towards the concluding stages of the evaporation.  

\section{Conclusions}\label{sec:conclusions}

In conclusion, quantum effects influence every stage of an RBH's lifetime. We demonstrated the existence of peculiar features such as the absence of the Hawking--Page transition, the deviation from the mean-field theory ratio at the critical point, and the different responses of a scalar field's entanglement entropy based on the regularization method employed. Additionally, assuming semiclassical gravity holds near the apparent horizons of an RBH, we demonstrated that the NEC is violated near the outer apparent horizon and satisfied near the inner horizon throughout its lifetime. This is inherently linked to the undesirable feature of the violation of the third law of black hole mechanics. Consequently, the regularization of spacetime, while desirable, unveils peculiar properties and raises unanswered questions that may only find resolution through a comprehensive theory of quantum gravity. Even when quantum effects are considered, the challenges are simply shifted to other areas, highlighting the complexities that require further understanding.

\section*{Acknowledgments}
I would like to thank Sebastian Murk, Fil Simovic, and Daniel Terno for useful discussions and helpful comments. I.S. is supported by an International Macquarie University Research Excellence Scholarship.

\end{document}